\documentclass[reprint, aps, pra, superscriptaddress]{revtex4-1}

\usepackage[latin1]{inputenc}
\usepackage[T1]{fontenc}
\usepackage{gensymb} 
\usepackage[seperr,load={}]{siunitx}
\usepackage{lmodern}
\usepackage{amsmath}
\usepackage{amssymb}
\usepackage{graphicx}
\usepackage{color}
\usepackage{soul}

\begin{document}

\title{Non-galvanic primary thermometry of a two-dimensional electron gas}
\author{P. Torresani}
\affiliation{SPSMS, CEA-INAC/UJF-Grenoble 1, 17 Rue des Martyrs, F-38054 Grenoble Cedex 9, France}
\author{M. J. Mart\'{i}nez-P\'{e}rez}
\affiliation{NEST, Istituto Nanoscienze-CNR and Scuola Normale Superiore, I-56127 Pisa, Italy}
\author{S. Gasparinetti}
\affiliation{Low Temperature Laboratory, Aalto University, P.O. Box 15100, FI-00076 Aalto, Finland}
\author{J. Renard}
\affiliation{SPSMS, CEA-INAC/UJF-Grenoble 1, 17 Rue des Martyrs, F-38054 Grenoble Cedex 9, France}
\author{G. Biasiol}
\affiliation{CNR-IOM, Laboratorio TASC, Area Science Park, I-34149 Trieste, Italy}
\author{L. Sorba}
\affiliation{NEST Istituto Nanoscienze-CNR and Scuola Normale Superiore, I-56127 Pisa, Italy}
\author{F. Giazotto}
\affiliation{NEST Istituto Nanoscienze-CNR and Scuola Normale Superiore, I-56127 Pisa, Italy}
\author{S. De Franceschi}
\affiliation{SPSMS, CEA-INAC/UJF-Grenoble 1, 17 Rue des Martyrs, F-38054 Grenoble Cedex 9, France}

\begin{abstract}
We report the experimental realization of a non-galvanic, primary thermometer capable of measuring the electron temperature of a two-dimensional electron gas with negligible thermal load. Such a thermometer consists of a quantum dot whose temperature-dependent, single-electron transitions are detected by means of a quantum-point-contact electrometer. Its operating principle is demonstrated for a wide range of electron temperatures from 40 to 800 mK. This noninvasive thermometry can find application in experiments addressing the thermal properties of micrometer-scale mesoscopic electron systems, where heating or cooling electrons requires relatively low thermal budgets. 

\end{abstract}

\maketitle

Reaching the millikelvin temperatures in low-dimensional systems has led to the discovery of a number of important phenomena, starting from the quantum Hall effect \cite{JPSJ.39.279}.  Access to even lower electron temperatures, i.e., below 1 mK, would allow a deeper inspection of the rich underlying physics and, possibly, open the door to a new horizon of unexplored phenomena (e.g., nuclear-spin ferromagnetic phase transitions \cite{simon2007nuclear}, topologically protected excitations in the $\nu$=5/2 fractional state \cite{willett2013quantum} or exotic Kondo effects \cite{potok2007observation}). Due to the suppression of electron-phonon coupling at low temperature, cooling of electrons by means of all-electrical methods seems more promising than cooling them indirectly via the crystal lattice \cite{clark:103904}. Hence, increasing efforts have been focusing on the development of electronic coolers in both semiconductor \cite{edwards:1815,PhysRevLett.102.146602} and metal systems \cite{PhysRevLett.92.056804,giazotto:3784,quaranta:032501,lowell:082601,muhonen2012micrometre}. In order to reach ultra-low temperatures, any potential source of electron heating needs to be minimized. Even the mere fact of measuring an electron temperature can add a significant power load and lead to appreciable heating. This problem 
emerges already below 100 mK for relatively small (micrometer-scale) electron systems, and it 
becomes particularly critical below ~1 mK, where measuring low electron temperatures in a non-invasive way can be as challenging as achieving them \cite{6176123}. 

In this paper, we report an experimental demonstration of non-galvanic, primary thermometry of a 2DEG in the $40-800$ mK temperature range. Here non-galvanic refers to the absence of a direct electrical connection between the measurement circuit and the 2DEG domain to be measured, and primary refers to the fact that no calibration against an absolute thermometer is required \cite{RevModPhys.78.217}. This thermometry concept, originally proposed in Ref. \cite{Prance_thesis} and theoretically analyzed in Ref. \cite{gasparinetti:253502}, is based on a quantum dot (QD) tunnel coupled to the 2DEG domain, and a nearby quantum point contact (QPC), used as a sensitive electrometer. 
The thermally broadened (hence temperature dependent) single-electron transitions in the QD charge are probed by means of the QPC, whose conductance is sensitive to the charge occupation of the QD.

We shall first validate this charge-detection (CD) thermometry scheme by comparing its results to those obtained by an already established approach based on Coulomb-blockade (CB) transport across the QD. 
Both of these methods, require the life-time broadening, $\Gamma$, of the QD levels to be much smaller than the thermal energy, $k_B T_e$, where $k_B$ is the Boltzmann constant and $T_e$ the electron temperature \cite{PhysRevB.44.1646}. Under this condition, the filling of a QD level swept through the Fermi energy of a 2DEG domain will occur according to the profile of the Fermi distribution function. In particular, the mean occupation of the QD will vary by one electron charge, $e$,  over an energy range set by $k_B T_e$. 

In the CD method, this charge information and the corresponding $T_e$ are read out in a non-galvanic fashion by means of the QPC 
\cite{PhysRevLett.70.1311,dicarlo2004differential,Gustavsson2009191}. In the CB method, the thermal broadening of a single-electron transition is measured by the width of the corresponding CB peak in the QD linear conductance.  
While in the latter case, the minimum life-time broadening required for a measurable current transport sets a lower bound, $k_B T_e \sim \Gamma$, on the operational temperature \cite{PhysRevB.83.201306}, here we shall show that the CD thermometry can be used also when $\Gamma$ is made so small that current transport across the QD is totally suppressed. Our results highlight the potential of CD thermometry to perform well even at temperatures much lower than those explored in this work, where a non-galvanic approach becomes necessary to ensure a reliable, non-invasive measurement of electron temperatures. 

The device was fabricated from a GaAs/AlGaAs heterostructure hosting a
100-nm-deep 2DEG of \SI{1.87e11}{cm^{-2}} charge density and \SI{1.26e6}{cm^2/Vs} mobility measured at 4.2 K in the dark.
A mesa structure was first defined by optical lithography and wet-etching in a $\rm{H_3PO_4}$ -- $\rm{H_2O_2}$
solution. The 2DEG was then contacted by means of annealed Au/Ge/Ni ohmic contacts.
Finally, surface split gates were patterned by e-beam lithography and Al deposition.
Most of the measurements were performed in a $^3$He cryostat with a base temperature of 260 mK. The measurement wiring consisted of, room-temperature pi-filters, constantan twisted pairs down to the $^3$He pot, followed by low-temperature RC filters with a cut-off frequency of 10 MHz and a resistance of 20 k$\Omega$. 
In order to test CD thermometry at and even lower temperatures, we used a cryo-free dilution refrigerator with a base temperature of 8 mK (in this case, copper-powder filters were incorporated in the measurement wiring at the level of the mixing chamber). 

\begin{figure}
\begin{center}
\includegraphics{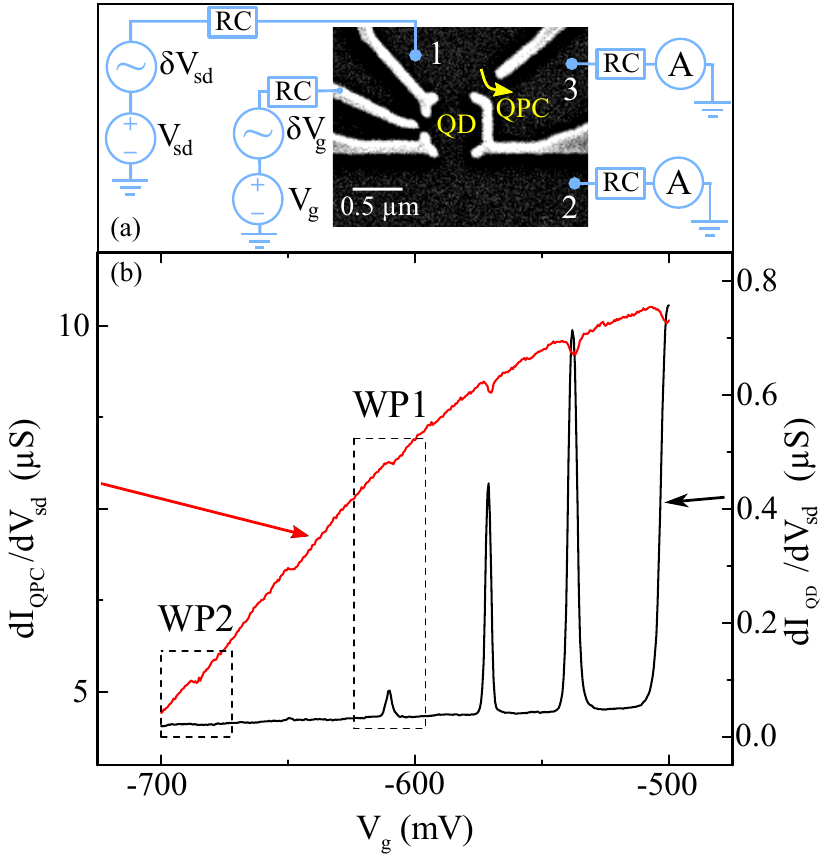}
\end{center}
\caption{\label{fig:prelim} (a) Scanning-electron micrograph of the device along with the measurement setup scheme. Surface gates define a QD and a QPC. Ohmic contacts are numbered. RC indicates low pass filters.
(b) QPC transconductance (red) and QD conductance (black) versus $V_g$. A \SI{20}{\micro V} ac bias is applied to contact 1. The two dashed boxes represent the two working points used during the measurements.}
\end{figure}

By applying a negative voltage to the split gates, the 2DEG is locally depleted such that 
a laterally confined QD and a QPC can be defined [see Fig.~\ref{fig:prelim}(a)]. In all measurements a bias voltage, $V_{sd}$, is applied to contact 1 (used as the source contact for both the QD and the QPC). Drain currents, $I_{QD}$ and $I_{QPC}$, are measured from contact 2 and contact 3, respectively. Linear conductance measurements are performed by low-frequency lock-in detection using an ac bias voltage $\delta V_{sd}$ smaller than $k_B T_e/e$. The QPC linear conductance $G_{QPC}$ exhibits the typical plateaus (not shown) characteristic of one-dimensional transport. Between two plateaus, $G_{QPC}$ is sensitive to the local electrostatic potential, such that the QPC can be used as a charge detector to probe the QD occupation. For our experiment, we tuned the QPC to the point of maximal slope (i.e., maximal charge sensitivity) below the first conductance plateau, corresponding to $G_{QPC} \sim e^2/h$, where $h$ is the Planck constant. 

We shall present first a complete set of data taken in the $^3$He setup. 
The QD was tuned into the CB regime denoted by the appearance of characteristic peaks in the QD linear conductance $G_{QD}$. A set of such Coulomb peaks is shown in Fig.~\ref{fig:prelim}(b) where $G_{QD}$ is plotted as a function of the voltage, $V_g$, applied to the plunger gate (black trace). Between consecutive peaks the QD holds an integer number of electrons and transport is blocked. From a finite bias measurement, $I_{QD}$ vs ($V_{sd}$, $V_g$), of the corresponding Coulomb diamonds (not shown), we estimate a QD charging energy, $U \sim 1$ meV and a level spacing, $\Delta \epsilon$, of a few hundred $\mu$eV.   
Figure ~\ref{fig:prelim}(b) also shows a simultaneous measurement of the QPC conductance (red trace). 
On top of a monotonic trend due to the capacitive coupling of the QPC to the QD plunger gate, the $G_{QPC}(V_g)$ characteristic exhibits a set of saw-tooth features in correspondence of the Coulomb peaks. These features stem from single-electron changes in the occupation of the QD. On the right-hand side of Fig. ~\ref{fig:prelim}(b), following the increased height of the Coulomb peaks in $G_{QD}$, the saw-tooth structures evolve into clear dips. This behavior is a voltage-division effect due to the common source contact between the QPC and the QD. Of the same origin is the small finite slope in the $G_{QD}(V_g)$ characteristic. Both these anomalies disappear 
when $G_{QD}(V_g)$ and $G_{QPC}(V_g)$ are independently measured one
after the other with the unused drain left floating.



For highly negative values of $V_g$ [left-hand side of Fig.~\ref{fig:prelim} (b)], the tunnel coupling between the QD and its leads becomes too small to allow for any measurable current through the QD. While Coulomb peaks get entirely suppressed, however, changes in the occupation number of the QD can still be detected by monitoring the QPC signal, which is the key for non-galvanic thermometry.  


Let us first demonstrate the equivalence between CD thermometry and the already established CB thermometry. To this aim, we chose a working point, labeled as WP1 in Fig.~\ref{fig:prelim} (b), where both a plateau in $G_{QPC}$ and a small, thermally-broadened Coulomb peak in $G_{QD}$ are simultaneously visible. 
In the conventional approach, $T_e$  is measured from the full-width at half maximum, $w$, of the Coulomb peak. This requires $U$, $\Delta \epsilon$, $\Gamma >> k_B T_e$. This condition is fulfilled by the Coulomb peak at WP1, whose $G_{QD}(V_g)$ profile is thermally broadened following the functional dependence \cite{kouwenhoven1997electron,PhysRevB.44.1646}: 


\begin{equation}
y = \bar{y} \cosh^{-2}\left[\frac{2 \ln{(1 +
\sqrt{2})} \times (x - \bar{x})}{w}\right]
\end{equation}
where $\bar{x}$ and $\bar{y}$ are the $V_g$-position and the $G_{QD}$-height of the Coulomb peak, respectively. It can be shown that $w = (3.52 k_B T_e)/\alpha$, 
where $\alpha$ is the lever-arm parameter of the plunger gate, i.e., the proportionality coefficient between a $V_g$ variation and the corresponding shift in the energy of the QD levels. 
The electron temperature is thus extracted from $w$ after fitting the Coulomb-peak profile to Eq. (1). 
An independent measurement of $\alpha$ is thus required. In general, $\alpha$ is related to the size and shape of the QD. Therefore,  $\alpha$ can vary significantly with the number of confined electrons. The value of $\alpha$ at the working point WP1 was obtained through a standard procedure \cite{kouwenhoven1997electron} from a measurement of the QD differential conductance, $dI_{QD}/dV_g$, as a function of ($V_g$, $V_{sd}$) [see upper inset of Fig. 2]. 
We found $\alpha = (4.17 \pm 0.2) \times 10^{-2}$ meV/mV.

In the CD approach, $T_e$ can be obtained by fitting the saw-tooth $G_{QPC}(V_g)$ profile or, alternatively, the corresponding dip in the transconductance 
$dI_{QPC}/dV_g$. It can be shown \cite{gasparinetti:253502} that the dip in $dI_{QPC}/dV_g$ has the same functional dependence as the Coulomb peak in $G_{QD}(V_g)$, the only difference being in the sign of $\bar{y}$. Therefore $T_e$ can be derived from the best-fit value of $w$ using the same $\alpha$ parameter. 

In order to show the good agreement between the two thermometry approaches, we plot in the lower inset of Fig. 2 a representative Coulomb peak (green trace) and the corresponding  d$I_{QPC}$/d$V_g$ dip (red trace), where the latter has been vertically rescaled and offset in order to achieve the best overlap. These measurements were  taken at the same fridge temperature, $T_f = 260$ mK. The almost perfect match between the two traces confirms that they indeed obey the same functional dependence outlined in Eq. (1). Moreover, since no rescaling along the horizontal axis was applied, the two traces yield consistent electron temperatures.    

Figure 2 shows two data sets of $T_e$ values measured at different $T_{f}$ using the CB and the CD methods, respectively.
For the QD-transport measurements, we applied a small ac bias voltage $\delta V_{sd}$ to contact 1 ($e \delta V_{sd} \approx  k_B T_e$) and measured the resulting ac current from contact 2, while keeping contact 3 open. For the CD method, we added a small ac modulation $\delta V_g$ to the dc plunger gate 
voltage ($\alpha \delta V_{g} \approx  k_B T_e$) and measured the resulting ac component of the QPC current, while keeping contact 2 open. 


The two data sets exhibit a good match over most of the temperature range. This establishes the equivalence between the two thermometry methods. The agreement between the two approaches becomes less good at the highest fridge temperatures, where adjacent Coulomb peaks begin to overlap each other affecting the reliability of the data fitting. 
We also note that $T_e$ closely follows $T_{f}$ (plotted as a reference dashed line) down to the lowest temperature. This is an indication of good thermalization and noise filtering of the measurement wiring. Simultaneously, the consistency between $T_e$ and $T_{f}$ demonstrates the fidelity of our thermometry methods highlighting their primary nature.


\begin{figure}
\begin{center}
\includegraphics{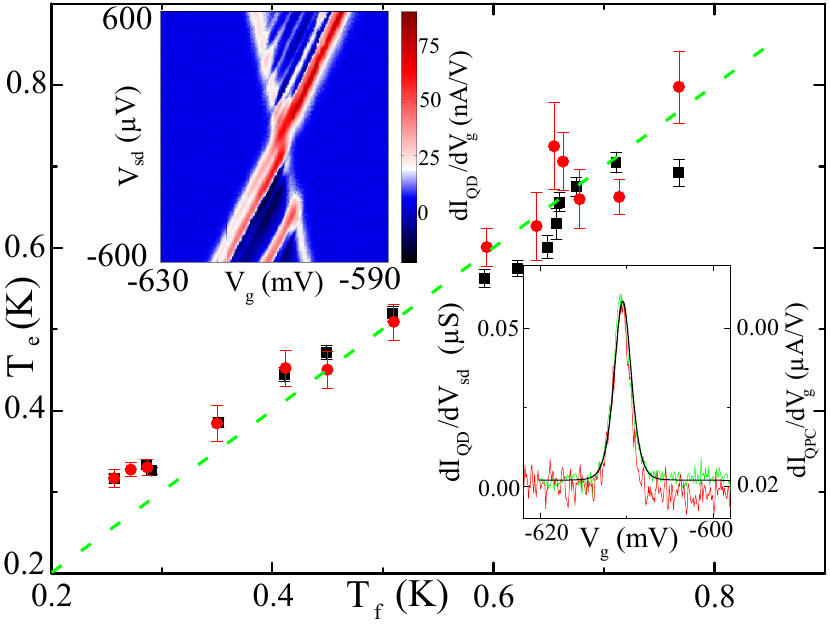}
\end{center}
\caption{\label{fig:galv_nongalv} Electron temperature $T_e$ versus fridge temperature $T_f$. $T_e$ is extracted from a CB measurement (black squares) and a CD measurement (red dots) at WP1 [see Fig.~\ref{fig:prelim}]. The error bars correspond to the fit error; an additional 5\% uncertainty comes from the measurement of the lever arm. The green dashed line represents $T_e = T_f$.
Top Left: finite-bias spectroscopy of the QD around WP1: QD conductance $G_{\rm QD}$ versus dc bias $V_{sd}$ and gate voltage $V_g$.
Bottom Right: Sample CB and CD measurement traces at base temperature ($T_f$ = 260 mK) and a fit of the CB trace (black). CB trace: $G_{\rm QD}$ versus $V_g$ (green). CD trace: gate-to-QPC transconductance versus $V_g$ (red).
The CD trace is horizontally offset.}
\end{figure}

\begin{figure}
\begin{center}
\includegraphics{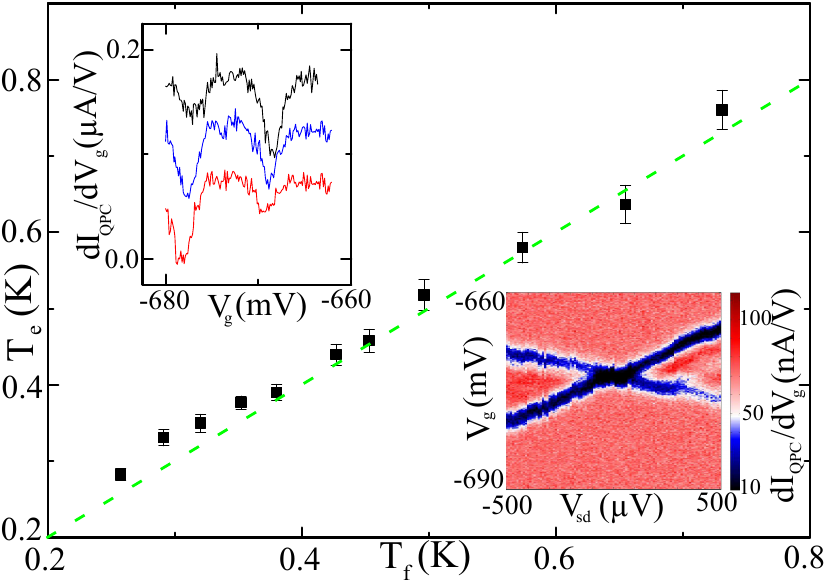}
\end{center}
\caption{\label{fig:deep_nongalv}
Electron temperature $T_e$ as extracted from a CD measurement at WP2 [see Fig.~1], versus fridge temperature $T_f$.
The error bars are given as in Fig.~\ref{fig:galv_nongalv}. The green dashed line represents $T_e = T_f$.
Top Left: gate-to-QPC transconductance at WP2 versus gate voltage $V_g$ for three different configurations of the dot barriers:
the lower barrier is less transparent (black), the two barriers have same transparency (blue) and the upper barrier is less transparent (red).
Bottom Right: gate-to-QPC transconductance at WP2 versus gate voltage $V_g$ and dc bias $V_{\rm sd}$. This plot is used to extract the lever arm at WP2.
}
\end{figure}

\begin{figure}
\begin{center}
\includegraphics{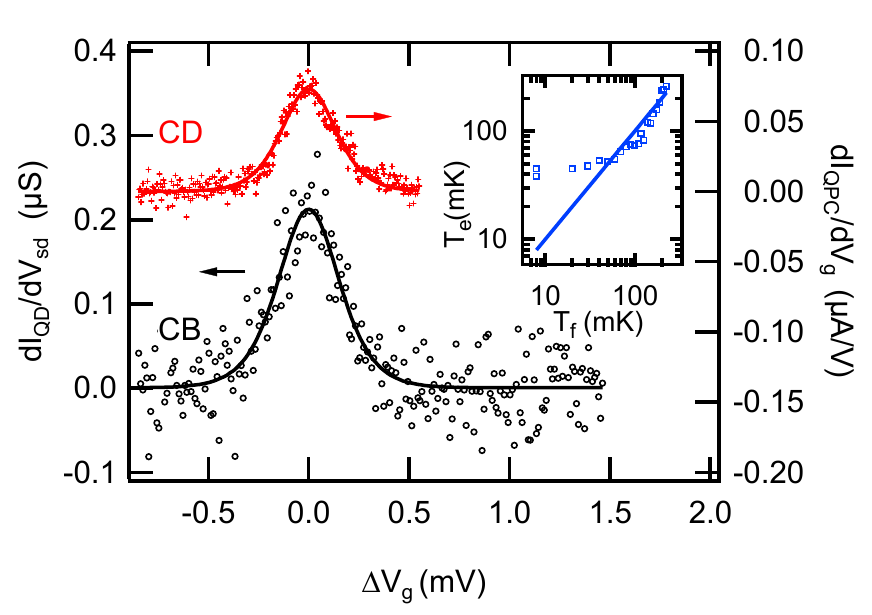}
\end{center}
\caption{\label{fig:figure4}
Comparison of CD (red crosses, right axis) and CB (black open circles, left axis) thermometry at a dilution-fridge base temperature of 8 mK. The bottom axis corresponds to the gate voltage relative to the peak position. The solid lines are fits to the Eq. (1) and give electronic temperatures $T_e = 42 \pm 3.5$ mK (CD) and $T_e = 46 \pm 4.5$ (CB). Inset: Electronic temperature measured using CB thermometry as a function of the fridge temperature, $T_f$. The solid line corresponds to  $T_e = T_f$.
}
\end{figure}

The second important step of this work is to demonstrate the applicability of the CD thermometry method to
the very weak coupling regime, where current transport through the QD is suppressed to well below the current detection limit (a few tens fA).
In addition, we want to show that it is possible to use the CD method to measure the electron distribution function of just one 2DEG domain, i.e., the drain reservoir.   
To this aim, we moved to the working point WP2 on the left-hand side of Fig. 1. Here only a saw-tooth structure in $G_{QPC}$ or, equivalently, a dip in
$dI_{QPC}/dV_g$ can be seen. 

An accurate estimate of $T_e$ from the dip profile requires a determination of the $\alpha$ parameter in this new regime.  Since no QD current is observed, however, an alternative method entirely based on CD is required. To address this problem, let us consider first the case where 
the tunnel couplings of QD to the source and drain leads ($\Gamma_s$ and $\Gamma_d$, respectively) are very low but equal. In this case, 
for a finite dc bias voltage across the QD, the average occupation of the QD is a half integer each time the QD electrochemical potential, $\mu_{QD}$, lies within the bias window (strickly speaking, this is true only for spinless electrons, which we have implicitly assumed here for the sake of simplicity; taking spin degeneracy into account the average charge on the dot would be $2e/3$ instead of $e/2$).
Therefore a $V_g$ sweep moving $\mu_{QD}$ accross the bias window will produce two dips in d$I_{QPC}$/d$V_g$, 
one for $\mu_{QD} = \mu_s$ and one for $\mu_{QD} = \mu_d$, where $\mu_s$ and $\mu_d$ are the Fermi energies of the source and drain leads, respectively.
The two dips should be identical as they both result from the addition of an average $e/2$ to the QD.  This is no longer the case when $\Gamma_s \neq \Gamma_d$, where an asymmetry in the dips is expected to arise. In particular,  when $\Gamma_s > \Gamma_d$ ($\Gamma_s < \Gamma_d$) the dip corresponding to $\mu_{QD} = \mu_s$ ($\mu_{QD} = \mu_d$) becomes dominant at the expense of the other dip. This effect is shown in the upper inset of Fig. 3 where three d$I_{QPC}$/d$V_g(V_g)$ traces are shown for the same $V_{sd}$ but different symmetry conditions:  $\Gamma_s > \Gamma_d$,  $\Gamma_s \approx \Gamma_d$, and $\Gamma_s < \Gamma_d$. 
The $V_g$ separation between the two dips is proportional to $V_{sd}$ enabling a measurement of the $\alpha$ parameter. 

The lower inset of Fig. 3 shows a color plot 
of d$I_{QPC}$/d$V_g$ as a function of ($V_g$,$V_{sd}$), which is analogous to the Coulomb diamond in the upper inset of Fig. 2. This data set was taken with $\Gamma_s < \Gamma_d$, i.e. in an asymmetric condition as close as possible to the limit $\Gamma_s << \Gamma_d$. In fact, it is in this limit that the QD is only sensitive to the thermal broadening of the electron distribution in the drain reservoir, which is our subject of interest. 
From the data in the lower inset of Fig. 3, we estimate $\alpha = (4.45 \pm 0.2) \times 10^{-2}$ meV/mV.  This value is 6.5\% larger than the previous one measured at WP1, i.e. just 80 mV higher in $V_g$ and for only two additional electrons on the QD. This relevant difference tells that, in order to avoid a systematic scaling error, it is important to measure the lever arm in the fully closed, non-galvanic regime, as close as possible to the conditions under which the CD thermometry is performed. 

Such conditions were achieved by further increasing the barrier asymmetry up to the point were only the d$I_{QPC}$/d$V_g$ dip associated with the drain reservoir is visible. Given the smallness of the gate-voltage adjustments (just a few mV) necessary to achieve this additional increase in the barrier asymmetry, we can assume the obtained $\alpha$ value to be appropriate for our thermometry measurements. The results of these measurements are shown in Fig. 3. As in the previous case of Fig. 2, the electron temperature (which now refers to the drain reservoir only) follows fairly well the fridge temperature. 

In order to perform a test of CD thermometry at even lower temperature, the same device was transferred to a dilution refrigerator. 
Figure 4 shows two measurements taken at $T_f = 8$ mK: a CD measurement of d$I_{QPC}$/d$V_g$ vs $V_g$ in the fully blocked, asymmetric regime (red data points); and a CB measurement of d$I_{QD}$/d$V_{sd}$ vs $V_g$ in a conductive regime (black data points). Numerical fits to Eq. (1) (solid lines) yield $T_e = 42 \pm 3.5$ mK and $T_e = 46 \pm 4.5$ mK, respectively. The given uncertainties include a systematic 5\% error on the $\alpha$ factor and the error on the numerical fit, which can be taken as a measure of the thermometer sensitivity. This experimental sensitivity (1.5 mK and 2 mK, for the CD and the CB data, respectively) approaches our expectation (~1 mK) based uniquely on the output noise level of the gate-voltage sources. The fact that $T_e$ is significantly larger than $T_f$ is due to an non-perfect filtering of the electrical noise reaching the sample through the measurement leads. The low-temperature saturation of $T_e$ is clearly shown in the inset to Fig. 4 (data from CB thermometry).  

The data of Fig. 4 show that, within experimental uncertainty, CD and CB thermometry's yield consistent results down to very low temperatures. This is not surprising since 40 mK is still well above the lowest temperature measurable by means of CB thermometry, which we estimate to be $\sim0.1$ mK (this estimate is obtained by imposing a minimum CB peak current of $\sim1$ pA). Such a low threshold temperature holds only in the absence of heating due to the current flow through the QD detector. This is true for large electron reservoirs, which is the case here, but not for mesoscopic reservoirs with lateral dimensions in the micrometer scale. In the latter case, it was shown that, because of the heat load introduced by the CB thermometer, $T_e$ cannot be lowered below $\sim100$ mK no matter how cold the lattice gets. Hence the use of CD thermometry was suggested as a potential solution to this heating problem. By demonstrating CD thermometry down to at least $T_e \approx 40$ mK, we have proved this solution to be experimentally viable and applicable to a relatively wide and accessible temperature range. 

In conclusion, we experimentally tested the operating principle of the non-galvanic CD thermometer discussed in Gasparinetti {\it et al.} \cite{gasparinetti:253502}. First, we demonstrated the equivalence between CD thermometry and the already established CB thermometry in a galvanic regime where a small, yet measurable current is driven through the probed electron reservoir. Then, we showed that the operation of the CD thermometer can be extended to the non-galvanic regime where current transport is suppressed preventing CB thermometry. 
The feasibility of CD thermometry was demonstrated down to an electron temperature of 40 mK.  
This opens concrete opportunities for experiments addressing the low-temperature thermal properties of mesoscopic electron reservoirs, including experiments on electron cooling. More precisely, for electron domains with lateral sizes in the micrometer range, CD thermometry can be an asset already at $T_e$ below ~100 mK, 
where CB thermometry becomes inadequate \cite{gasparinetti:253502}. 
Finally, we stressed the importance of performing an accurate measurement of the lever-arm parameter $\alpha$. This is a necessary requirement for CD and CB thermometers to  be used as primary thermometers. We showed that $\alpha$ can significantly vary when a QD is tuned from a conducting regime to pinch-off. In the latter case, which is relevant for non-galvanic thermometry, $\alpha$ cannot be measured from finite-bias CB transport as usual. Here we showed that a CD method can be used to measure $\alpha$. 

{\it \underline{Note added}}. Recently we have become aware of a similar work by Mavalankar {\it et al.} \cite{mavalankar}.

M.~J.~M.-P.~and F.~G.~acknowledge the FP7 program No.~228464 MICROKELVIN, the
Italian Ministry of Defense through the PNRM project TERASUPER, and the Marie
Curie Initial Training Action (ITN) Q-NET 264034 for partial financial support.
S.~G. acknowledges financial support from the Finnish National Graduate School in Nanoscience.
P. T., J. R., and S.D. acknowledge support from the EU through the ERC starting grant HybridNano.  

\bibliography{biblio}

\end{document}